\documentclass[
 reprint,
 amsmath,
 amssymb,
 aps,
 prd,
 preprintnumbers,
 nofootinbib
]{revtex4-2}

\usepackage{slashed}
\usepackage{amsmath}
\usepackage{graphicx}
\usepackage{xspace}

\usepackage{xcolor}
\definecolor{green}{HTML}{1b9e77}

\newcommand{\Z}{\mathbb{Z}}

\newcommand{\SU}{\mathrm{SU}}

\newcommand{\DCSB}{D$\chi$SB\xspace}
\begin{document}

\title{Numerical indication that center vortices drive dynamical mass generation in QCD}

\author{Waseem Kamleh}\email{Corresponding author [waseem.kamleh@adelaide.edu.au]} \author{Derek B. Leinweber} \author{Adam Virgili}
\affiliation{Special Research Centre for the Subatomic Structure of
 	Matter (CSSM), Department of Physics, University of Adelaide,
 	South Australia 5005, Australia.}

\begin{abstract}
The first calculation of the response of the momentum space quark propagator to center vortices in the ground state fields of QCD is presented. Center vortices are identified on 2+1-flavour dynamical gauge fields with $m_\pi \simeq 156$ MeV to obtain the vortex-removed and vortex-only quark propagator. Dynamical mass generation is found to vanish upon vortex removal, while the vortex-only field is able to generate dynamical mass. These new signatures strengthen the lattice QCD evidence indicating that center vortices underpin both dynamical chiral symmetry breaking and quark confinement.
\end{abstract}

\preprint{ADP-23-14/T1223}

\maketitle

Quark confinement and dynamical chiral symmetry breaking (\DCSB) are two emergent features of quantum chromodynamics (QCD) that are foundational to the structure of the visible Universe. The formation of the proton and neutron that are the building blocks of matter is a consequence of the confinement of quarks into hadrons. The nucleon mass is almost entirely generated dynamically by the strong interaction (noting that the Higgs mechanism only accounts for $\sim1\%$ of the nucleon rest energy). To date there has been no direct analytic derivation of either confinement or the dynamical mass generation associated with \DCSB from the QCD Lagrangian. Hence, the possibility of a single unified mechanism that underpins both quark confinement and dynamical chiral symmetry breaking is an enticing prospect. There is now a substantial body of calculations supporting the center vortex model as a strong candidate for this mechanism~\cite{tHooft:1977nqb,tHooft:1979rtg,DelDebbio:1996lih,Faber:1997rp,DelDebbio:1998luz,Engelhardt:1998wu,Langfeld:1998cz,Bertle:1999tw,deForcrand:1999our,Faber:1999gu,Engelhardt:1999fd,Alexandrou:1999iy,Alexandrou:1999vx,Engelhardt:1999xw,Engelhardt:1999wr,Engelhardt:2000wc,Bertle:2000qv,Bertle:2001xd,Langfeld:2001cz,Engelhardt:2002qs,Langfeld:2003ev,Greensite:2002yn,Greensite:2003bk,Greensite:2003xf,Bruckmann:2003yd,Engelhardt:2003wm,Boyko:2006ic,Greensite:2006ng,Ilgenfritz:2007ua,Bornyakov:2007fz,Bowman:2008qd,Hollwieser:2008tq,OCais:2008kqh,Engelhardt:2010ft,Bowman:2010zr,OMalley:2011aa,Hollwieser:2013xja,Hollwieser:2014soz,Trewartha:2015ida,Trewartha:2015nna,Greensite:2016pfc,Trewartha:2017ive,Biddle:2018dtc,Spengler:2018dxt,Biddle:2019gke,Biddle:2022zgw,Biddle:2022acd,Sale:2022qfn,Virgili:2022ybm,Biddle:2023lod}.

Lattice gauge theory provides a computational formalism that allows for the direct study of center vortex phenomenology, established by significant pioneering work~\cite{DelDebbio:1996lih,Faber:1997rp,DelDebbio:1998luz,Engelhardt:1998wu,Langfeld:1998cz,Bertle:1999tw,deForcrand:1999our,Faber:1999gu,Engelhardt:1999fd,Engelhardt:1999xw,Engelhardt:1999wr,Engelhardt:2000wc,Bertle:2000qv,Bertle:2001xd,Langfeld:2001cz,Engelhardt:2002qs,Langfeld:2003ev,Greensite:2002yn,Greensite:2003bk}. In a spacetime with $D$ dimensions, a thin center vortex is a closed $D-2$ dimensional hypersurface of center flux contained within the gauge manifold. In three dimensions center vortices are linelike, and in four dimensions they are surfacelike. On the lattice, the gluon field is described by a set of gauge links that belong to $\SU(3).$ Hence, the relevant center group is the set of elements that universally commute $\{ z I \} \simeq \Z_3,$ where $z$ is one of the three center phases,
\begin{equation}
z = \exp\left(\frac{2\pi i}{3}m\right), \quad m \in \left\{ -1,0,1 \right\}.
\end{equation}

A key advantage of the lattice prescription is the ability to construct \emph{vortex-modified} ensembles. The \emph{untouched} lattice gauge links $U_\mu(x)\in\SU(3)$ are decomposed into the following form
\begin{equation}
U_{\mu}(x) = Z_{\mu}(x)\, R_{\mu}(x),
\end{equation}
where the \emph{vortex-only} field $Z_{\mu}(x)\in\Z_3$ consists only of center elements, and the \emph{vortex-removed} field $R_{\mu}(x) = Z^*_{\mu}(x)\,U_{\mu}(x)$ describes the remaining gauge interactions.

The vortex-only field $Z_{\mu}(x)$ is defined by first fixing the gauge, typically to maximal center gauge~\cite{DelDebbio:1996lih,Langfeld:1997jx,Montero:1999by,Faber:1999sq,Langfeld:2003ev,OCais:2008kqh,Biddle:2022zgw}, and then projecting the links to the nearest center element. Vortices themselves exist on the dual lattice, and are identified with center-projected plaquettes that enclose a nontrivial center flux. These P-vortices correspond to \emph{thin} center vortices, describing the center flux along a two-dimensional surface. This surface would have some finite thickness for a physical center vortex. Although the identified thin vortices are gauge dependent, they do correlate with the physical thick vortices~\cite{DelDebbio:1998luz,Engelhardt:1999xw} and it is possible to learn about the latter by studying the former.
Numerous lattice studies comparing results from the untouched and vortex-modified ensembles have proven fruitful, and analysis of the differences that emerge from the absence or presence of center vortices grants insight into the role these structures play.

Reference~\cite{Biddle:2023lod} presents visualizations and analysis of the identified center vortex structure on representative configurations selected from the same dynamical QCD gauge field ensemble that we use herein. A key feature is the presence of a cluster of vortices that percolates throughout the lattice on each configuration. There are simple arguments showing that the percolation of center vortices through the four-dimensional lattice volume gives rise to an area law falloff for the Wilson loop, which then implies quark confinement~\cite{Engelhardt:1998wu,Greensite:2006ng}.

\begin{figure*}[t]
  {\includegraphics[width=8.6cm]{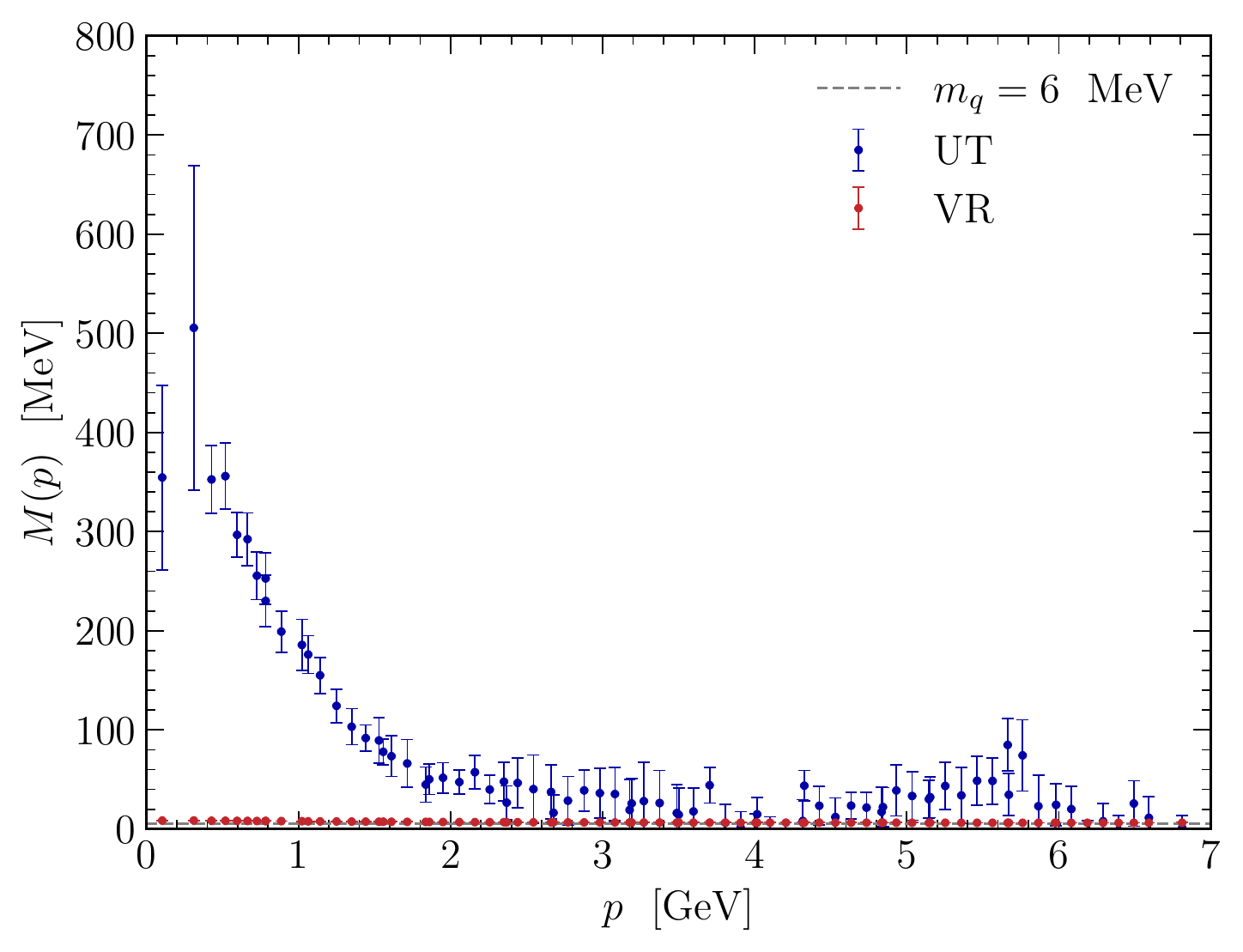}} 
  {\includegraphics[width=8.6cm]{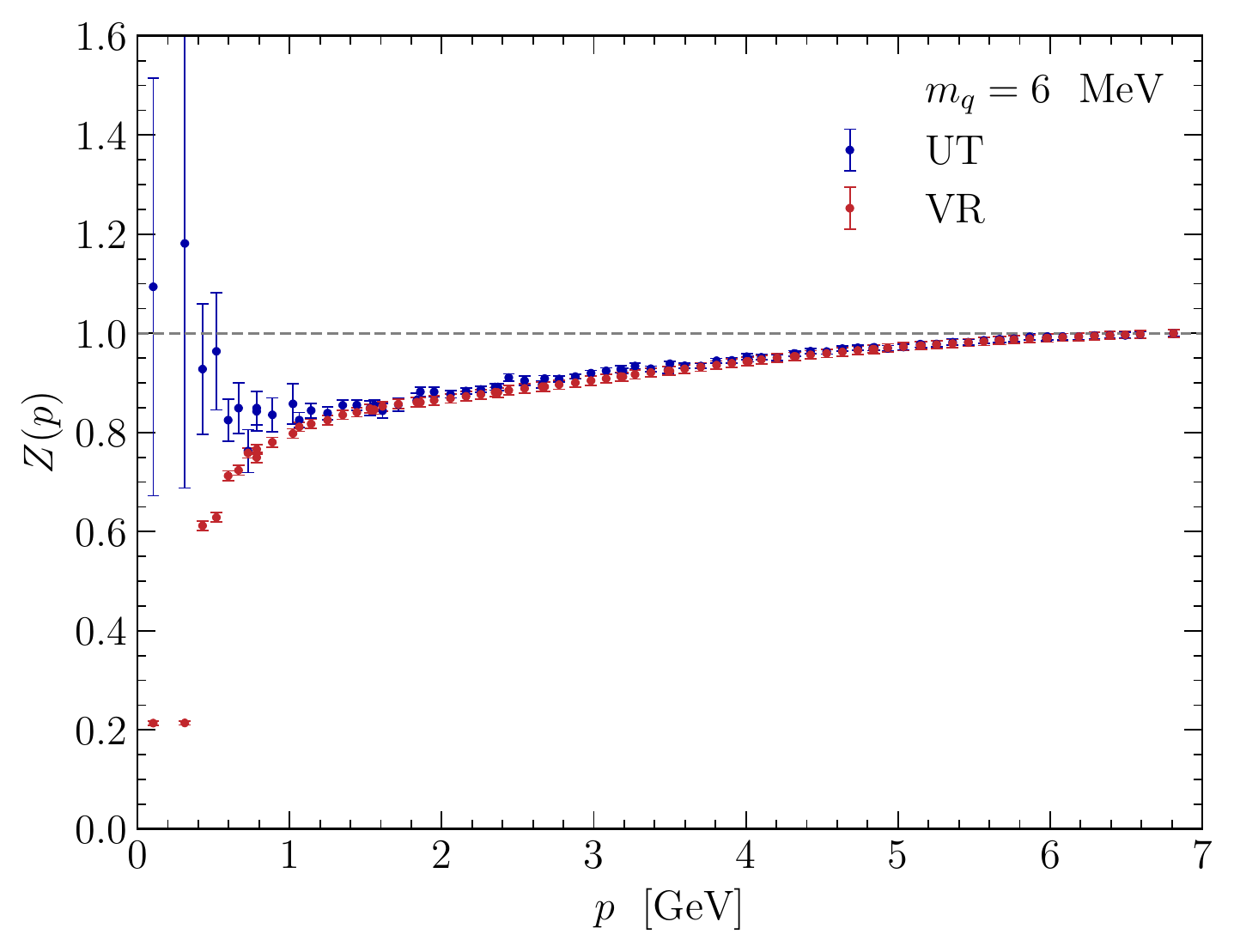}} 
  \caption{The quark mass function $M(p)$ (left) and the renormalized quark dressing function $Z(p)$ (right) for the untouched (UT) and vortex-removed (VR) ensembles. Results are calculated at $m_q=6$ MeV, with the dashed line indicating the respective tree-level values. The renormalization scale is set to $\nu = 6.8\text{ GeV},$ such that $Z(p)|_{p=\nu} = 1.$}
  \label{fig:UTVR}
\end{figure*}
In this Letter, we perform the first investigation of the connection between vortices and dynamical mass generation in full QCD, reporting results for the vortex-removed and vortex-only quark propagators calculated near the physical point. Our simulations are performed on 2+1 flavour gauge fields generated by the PACS-CS Collaboration~\cite{PACS-CS:2008bkb}. These $32^3 \times 64$  lattice configurations use a nonperturbatively improved clover fermion action for the dynamical up, down, and strange quarks, with the light quarks providing a pion mass $m_\pi = 156$ MeV close to the physical value and a lattice spacing $a=0.0933$ fm set by the S\"ommer parameter $r_0 = 0.49$ fm.

The valence quark propagator is calculated using the overlap fermion action due to its superior chiral properties~\cite{Narayanan:1993zzh,Narayanan:1993sk,Narayanan:1993ss,Narayanan:1994gw,Neuberger:1997fp,Kikukawa:1997qh}. 
The massive overlap Dirac operator is given by~\cite{Neuberger:1997bg}
\begin{equation}
D_{\rm o}(\mu) = (1-\mu)\,\frac{1}{2a}(1 + \gamma_5\,\epsilon[H(m_{\rm w})]) + \mu,
\end{equation}
where $\epsilon$ is the matrix sign function applied to the fermion matrix kernel $H(m_{\rm w}).$ The matrix kernel is typically chosen to be a Wilson-type fermion action or variant thereof~\cite{Kamleh:2001ff,Bietenholz:2002ks,Kovacs:2002nz,DeGrand:2004nq,Durr:2005mq,Durr:2005ik,Bietenholz:2006fj}, noting that a negative value for the Wilson mass parameter in the topological interval $0 < -am_{\rm w} < 2$ ensures the absence of fermion doublers. Here, we choose $am_{\rm w} = -1.1$ and $\mu = 0.0012,$ representing fermions with a bare quark mass $m_q = 2m_{\rm w}\mu \simeq 6\text{ MeV}.$ The external overlap quark propagator is given by~\cite{Narayanan:1994gw,Edwards:1998wx}
\begin{equation}
    S(p) \equiv \frac{1}{2m_{\rm w}(1-\mu)}(D^{-1}_o(\mu) - 1) \,,
\end{equation}
obeying a chiral symmetry $\{\gamma^5 ,S(p)\vert_{m_q = 0} \} = 0$ just as in the continuum~\cite{Neuberger:1997fp}.
Dynamical chiral symmetry breaking is linked to the dynamical generation of mass in the quark propagator~\cite{Fomin:1983kyk,Henley:1989vi,Williams:1989tv,Krein:1990sf,Alkofer:2000wg,Skullerud:2000un,Skullerud:2001aw,Bonnet:2002ih,Bowman:2002bm,Zhang:2004gv,Bowman:2005vx,Furui:2006ks,Kamleh:2007ud,Blossier:2010vt,Schrock:2011hq,Burger:2012ti,Pak:2015dxa,Oliveira:2018lln,Virgili:2022wfx}. Hence, the use of a valence quark action that has good chiral symmetry is advantageous.

The overlap quark propagator in momentum space takes the form
\begin{equation}
    S(p) = \frac{Z(p)}{i \slashed{q} + M(p)},
    \label{eq:Sprop}\end{equation}
where $M(p)$ is the quark mass function and $Z(p)$ is the quark dressing function. The only correction required for the overlap quark propagator is the identification of the kinematical lattice momentum $q_\mu,$ such that when the gauge links are set to unity the tree-level propagator takes the form 
\begin{equation}
    S(p)\big|_{U=1} = \frac{1}{i \slashed{q} + m_q}.
    \label{eq:Stree}
\end{equation}
To examine the structure of the quark propagator we fix to Landau gauge~\cite{Bonnet:1999mj,Hudspith:2014oja}. A momentum cylinder cut~\cite{Leinweber:1998im} is applied to minimize lattice artifacts.
The quark dressing function is renormalized using the MOM scheme~\cite{Leinweber:1998uu}, such that $Z(p)|_{p=\nu} = 1$ at the renormalization scale set by the largest available momenta $\nu=6.8$ GeV.

\begin{figure*}[t]
  {\includegraphics[width=8.6cm]{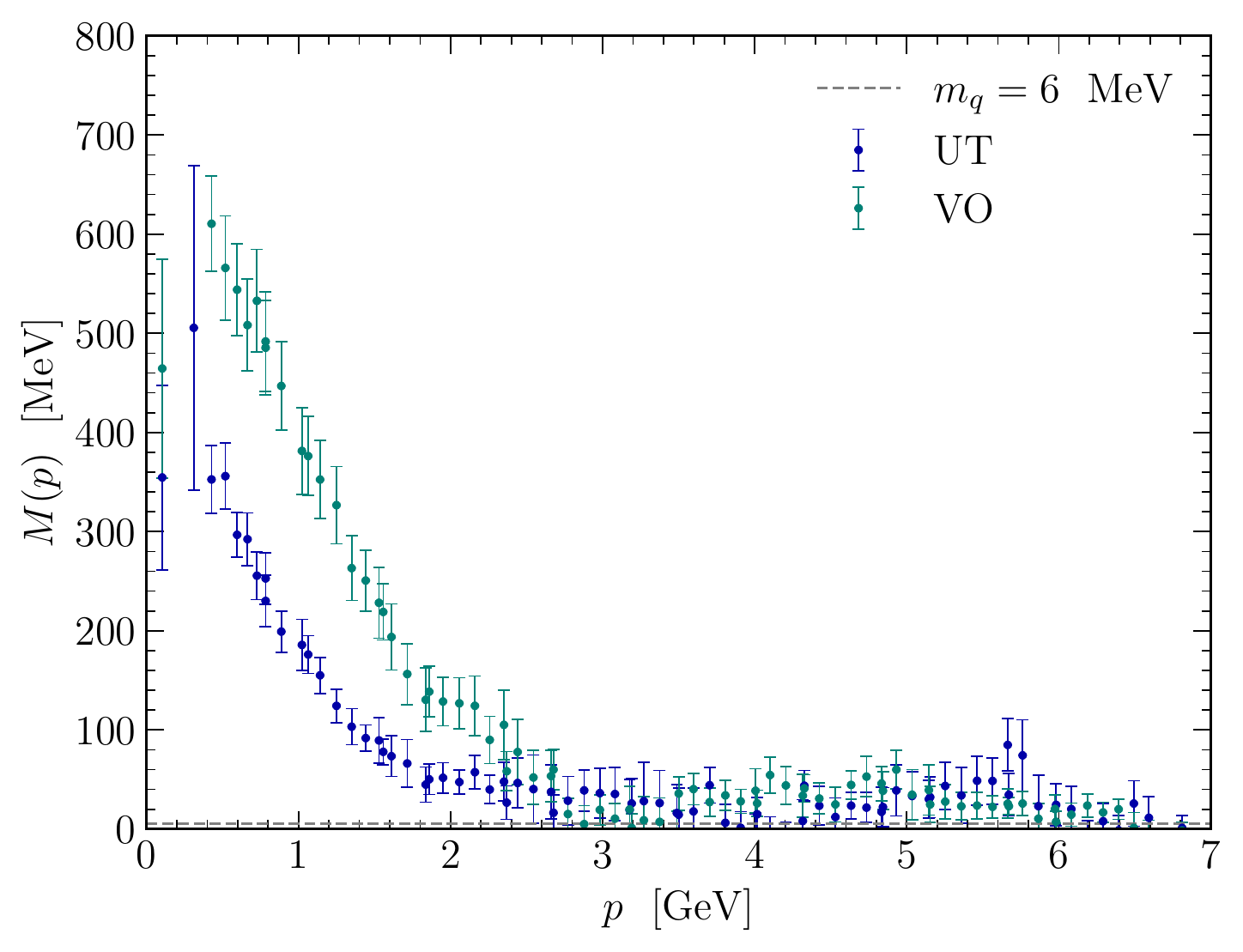}}
  {\includegraphics[width=8.6cm]{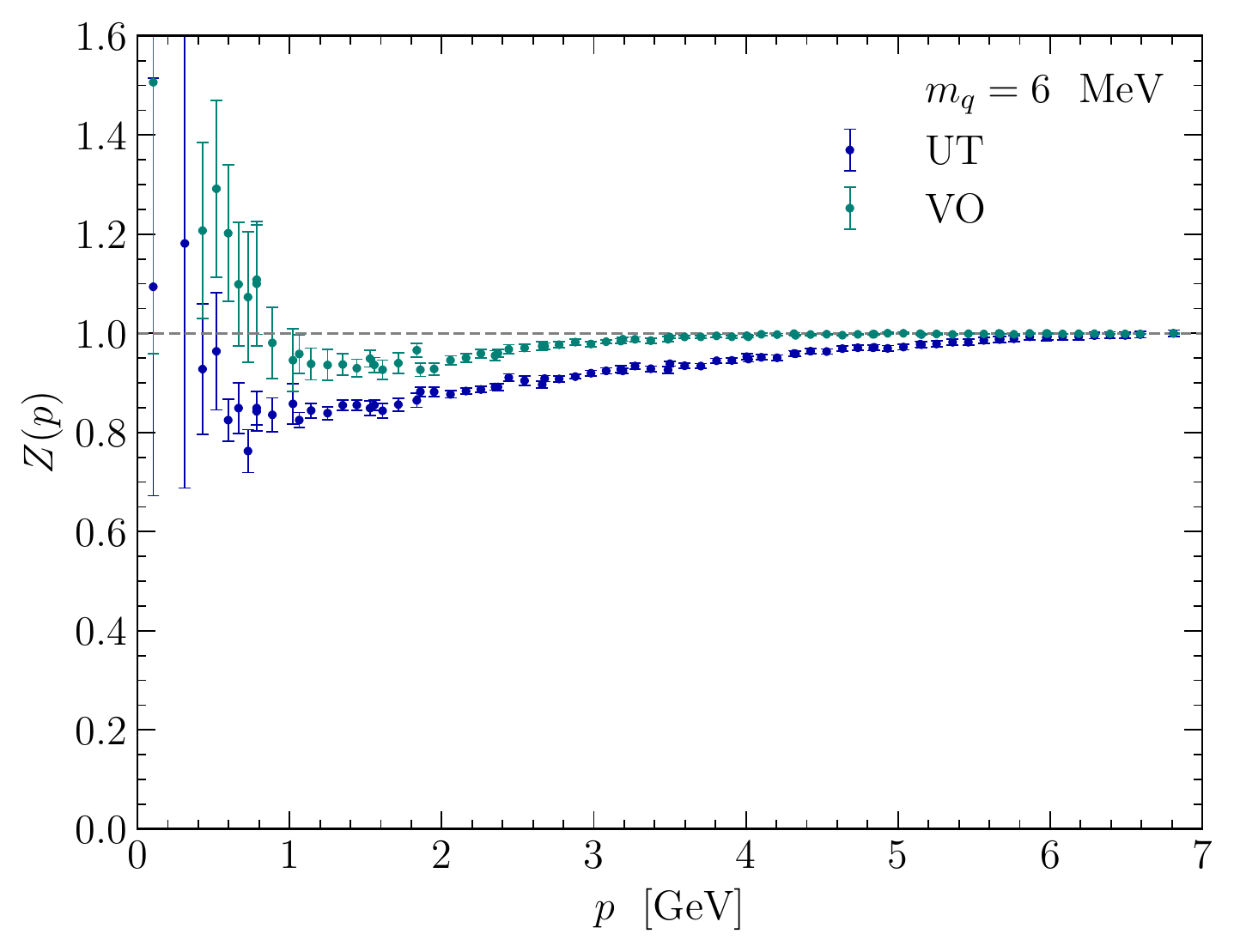}} 
  \caption{The quark mass function $M(p)$ (left) and the renormalized quark dressing function $Z(p)$ (right) for the untouched (UT) and vortex-only (VO) ensembles. Results are calculated at $m_q=6$ MeV, with the dashed line indicating the respective tree level values. The renormalization scale is set to $\nu = 6.8\text{ GeV}.$}
  \label{fig:UTVO}
\end{figure*}
We first consider the effect of vortex removal on the quark propagator. Our results for the vortex-removed propagator are calculated on 30 gauge field configurations, with all parameters matched to the untouched quark propagator results presented in Ref~\cite{Virgili:2022wfx}. In particular, both the vortex-removed and untouched results use a fat-link irrelevant clover (FLIC) fermion action~\cite{Zanotti:2001yb,Kamleh:2001ff,Kamleh:2004xk,Kamleh:2004aw} for the matrix kernel $H(m_{\rm w}),$ with four sweeps of stout-link smearing~\cite{Morningstar:2003gk} at $\rho=0.1$ applied to the irrelevant operators. The matrix sign function is evaluated using the Zolotarev rational polynomial approximation~\cite{Chiu:2002eh} with explicit projection of the 80 lowest-lying eigenmodes of the kernel.

Figure~\ref{fig:UTVR} shows the vortex-removed mass function and renormalization function, along with the corresponding untouched results from Ref.~\cite{Virgili:2022wfx}. The large $p$ behaviour of the untouched mass function is consistent with asymptotic freedom, with the approach to the tree-level value indicating that the coupling of the quark interactions to the gluonic vacuum diminishes in strength as $p$ increases. At small $p$ the key feature of the untouched results is the dynamical generation of mass, with indications of a peak or plateau at $\sim 350$ MeV, around one third of the nucleon mass as would be expected from constituent quark models. In comparison, the vortex-removed mass function is essentially identical to the tree-level value for the bare quark mass at all momenta. This is a truly remarkable result, indicating that the consequence of vortex removal is the \emph{elimination of dynamical mass generation.} This elimination of dynamical mass generation in QCD contrasts early results in the pure gauge sector, where $\sim 100$ MeV of mass generation would persist upon vortex removal~\cite{Trewartha:2015nna}. This new result establishes the relationship between center vortices and mass generation in QCD.

The untouched and vortex-removed quark dressing functions are broadly consistent for $p > 1\text{ Gev},$ but below this value they diverge. The untouched dressing function becomes noisier but is clearly approaching a finite value $\gtrsim 0.8$ as $p \to 0,$ whereas the vortex-removed dressing function tends toward zero. This behaviour can be understood as a consequence of the finiteness of $S(p)$ imposed by the lattice regularization. Specifically, at this very small bare quark mass the vortex removed mass function $M(p) \approx 0,$ and as $p \to 0$ we also have $q \to 0$, such that according to Eq.~(\ref{eq:Sprop}) we require $Z(p) \to 0$ as well in order for $S(p)$ to remain finite. 

Having seen that vortex removal eliminates dynamical mass generation, it is natural to examine the extent that the vortex-only fields are able to create it. This is somewhat complicated by the fact that it is not possible to perform these simulations directly on the vortex-only field. Unlike the more prevalent fermion discretizations, overlap fermions are not ultralocal due to the presence of the matrix sign function. Instead, overlap fermions satisfy an exponential locality that is conditional on the smoothness of the underlying gauge field~\cite{Hernandez:1998et,Adams:1999if,Neuberger:1999pz}. The vortex-only gauge field consisting only of center elements is too rough for the locality condition of the overlap fermion action to be satisfied, and hence some form of smoothing must be performed first.

The approach we take here is to first apply one pass of \emph{centrifuge preconditioning}~\cite{Virgili:2022ybm} at $\omega=0.02$, a process that perturbs the diagonal elements of the $\Z_3$ links away from the center group into three copies of $\mathrm{U}(1).$ In the spirit of gradient flow~\cite{Luscher:2009eq,Luscher:2010iy}, annealed smoothing via the overimproved AUS algorithm~\cite{Bonnet:2000dc} is applied for $N=1190$ sweeps at $\alpha=0.02$ in order to ensure the locality of the overlap action. In place of the FLIC kernel selected for the untouched and vortex removed results, the vortex-only calculation uses a clover action constructed only from the smoothed $Z_\mu(x)$ links. We use the same value of $m_{\rm w}$ for the clover and FLIC matrix kernels, such that the tree-level propagators are identical. Due to the high level of noise in the vortex-only results, we increase our statistics to 60 gauge field configurations. The number of eigenmodes explicitly treated in the matrix sign function is also increased to 150. All other simulation parameters remain the same as for the untouched calculation.

The vortex-only mass function and renormalization function are compared with the corresponding untouched results in Fig.~\ref{fig:UTVO}. Examining the mass function, we see that the smoothed vortex-only field is indeed able to create dynamical mass generation. The dynamical mass is actually in excess of the untouched result, with the vortex-only mass function peaking somewhere around 600 MeV, some 250 MeV above the untouched peak. For $p > 3$ GeV, the vortex-only and untouched mass functions are consistent.
The vortex-only quark dressing function shows similar qualitative behaviour to the untouched dressing function, but this does not extend to a quantitative level of agreement until we reach a very large value of $p$.

It is worthwhile to compare what we have observed here in dynamical QCD with previous studies in pure gauge theory. A number of results demonstrate a connection between center vortices in $\mathrm{SU}(2)$ with dynamical chiral symmetry breaking~\cite{deForcrand:1999our,Engelhardt:2002qs,Bornyakov:2007fz,Hollwieser:2008tq,Bowman:2008qd,Hollwieser:2013xja,Hollwieser:2014soz}. Indeed, a study of the staggered fermion quark propagator in $\mathrm{SU}(2)$ presented similar findings~\cite{Bowman:2008qd}, where dynamical mass generation is seen to vanish upon vortex removal at small quark mass, and vortex-only fields are seen to create dynamical mass generation.

As mentioned earlier, previous results for the vortex-modified overlap quark propagator in $\mathrm{SU}(3)$ gauge theory~\cite{Trewartha:2015nna} show that mass generation is diminished upon vortex removal, but at small bare quark mass there is still a significant residual dynamical mass in the vortex-removed mass function. The same pure gauge study showed that cooled vortex-only fields are able to quantitatively match the dynamical mass generation from cooled untouched fields. Understanding the influence of dynamical fermions on the vortex-modified propagator is the natural next step that we have addressed herein.

Dynamical mass generation is a clear signal of dynamical chiral symmetry breaking. In the context of $\SU(2)$ and $\SU(3)$ pure gauge theory, and here in full QCD with dynamical fermions, we see that dynamical mass generation is eliminated or reduced upon vortex removal, and is able to be created by vortex-only fields. The connection between vortices and \DCSB is strengthened by hadron spectrum results in the pure gauge theory, with a Wilson fermion calculation of the spectrum on vortex-removed ensembles displaying an absence of \DCSB~\cite{OMalley:2011aa}.

In another study with overlap fermions~\cite{Trewartha:2017ive}, at light quark masses the vortex-removed hadron spectrum is found to be consistent with the restoration of chiral symmetry. The pion is no longer a psuedo-Goldstone boson, and hadronic currents related by a chiral transformation (such as the $N$ and the $\Delta$) become degenerate. The same study showed that the vortex-only fields are able to reproduce the key features of the light hadron spectrum.

The combined weight of previous results with those presented in this Letter provides further numerical evidence that center vortices are the fundamental mechanism that is responsible for dynamical chiral symmetry breaking. The significance of center vortices is amplified when we review the results of other studies that provide similar links with the confinement of quarks.

The presence of a linear string tension in the static quark potential is fundamentally connected with the Wilson area law criterion for confinement. The center dominance picture for the asymptotic string tension has been explored in the context of $\SU(2)$ and $\SU(3)$ pure gauge theory~\cite{Langfeld:2003ev,OCais:2008kqh,Trewartha:2015ida}, and more recently in full QCD with dynamical fermions~\cite{Biddle:2022zgw}. In all three contexts, the string tension is observed to vanish on the vortex-removed fields. In all three contexts, the vortex-only fields generate a linear static quark potential.

The connection between center vortices and confinement is further strengthened by studies of the gluon propagator~\cite{Bowman:2010zr,Biddle:2018dtc,Biddle:2022acd}.
The link is clearest in an examination of the Euclidean time correlator~\cite{Biddle:2022acd}. The untouched and vortex-only correlators show violations of spectral positivity, indicating that gluons are confined in the context of $\SU(3)$ pure gauge theory and in full QCD. These violations are almost eliminated upon vortex removal in the pure gauge theory. In full QCD the vortex-removed correlator is non-negative, indicating the existence of a spectral representation for the gluon propagator and implying that gluons are deconfined.

This Letter adds to the extensive body of lattice QCD results indicating that center vortices provide a unified mechanism that underpins both confinement and \DCSB, a body which now encompasses dynamical QCD as well as pure gauge theory. This certainly marks a milestone for the center vortex model, but it is by no means the end of the story.

The extent to which it is reasonable to expect that a vortex-only field of center elements should be able to quantitatively recreate the untouched gauge field results is an open question, though some pieces of the puzzle are already present. In $\SU(2)$ the string tension generated by the vortex-only fields is able to recreate the full untouched string tension. However, in $\SU(3)$ gauge theory the vortex-only string tension has historically only been observed to be around sixty percent of the untouched value~\cite{OCais:2008kqh,Trewartha:2015ida,Langfeld:2003ev}. In full QCD with dynamical fermions, the vortex-only and untouched string tensions show agreement after novel modifications to the untouched potential fits are implemented in the Coulomb term~\cite{Biddle:2022zgw}.

In $\SU(2)$ gauge theory the vortex-only mass function calculated with staggered fermions is noisy, but lies well above the untouched mass function~\cite{Bowman:2008qd}. In $\SU(3)$ gauge theory, after some cooling is applied both the vortex-only and untouched mass functions calculated using overlap fermions show agreement~\cite{Trewartha:2015nna}. Here in dynamical QCD, we see the vortex-only mass function exceeds the untouched value in the infrared.

The precise meaning of quark confinement in the presence of matter fields has been investigated in the context of $\SU(2)$ gauge-Higgs theory~\cite{Fradkin:1978dv,Bertle:2003pj,Greensite:2003bk,Greensite:2006ng,Greensite:2017ajx,Greensite:2018mhh,Greensite:2020nhg}. As dynamical fermions allow for string breaking, extending these concepts to full QCD would be of interest. Indeed, significant differences have been observed in the identified center vortex structure in pure gauge theory~\cite{Biddle:2019gke} and full QCD~\cite{Biddle:2023lod}.

A deeper understanding of the response of center vortices to the presence of dynamical fermions will be the subject of future investigations. In particular, a calculation of the vortex-removed hadron spectrum in dynamical QCD would be an interesting follow-up to previous pure gauge studies~\cite{OMalley:2011aa,Trewartha:2017ive}.

The \textsc{cola} software library~\cite{Kamleh:2022nqr} was used to calculate the lattice results reported herein.

\begin{acknowledgments}
We thank the PACS-CS Collaboration for making their configurations available via the International Lattice Data Grid (ILDG).
This research was undertaken with resources provided by the Pawsey Supercomputing Centre through the National Computational Merit Allocation Scheme with funding from the Australian Government and the Government of Western Australia. Additional resources were provided from the National Computational Infrastructure (NCI) supported by the Australian Government through Grant No. LE190100021 via the University of Adelaide Partner Share.
This research is supported by Australian Research Council through Grants No. DP190102215 and No. DP210103706.
W.K. is supported by the Pawsey Supercomputing Centre through the Pawsey Centre for Extreme Scale Readiness (PaCER) program.
\end{acknowledgments}


%

\end{document}